# Sleptsov Nets are Turing-complete

**Bernard Berthomieu**

LAAS-CNRS
Université de Toulouse, CNRS,
Toulouse, France
7, avenue du Colonel Roche,
31031 Toulouse cedex 4, France
Bernard.Berthomieu@laas.fr

**Dmitry A. Zaitsev**

Department of Computer Science,
Darmstadt University of Technology,
Darmstadt, Germany
Karolinenplatz 5,
64289 Darmstadt, Germany
daze@acm.org

**Abstract.** The present paper proves that a Sleptsov net (SN) is Turing-complete, that considerably improves, with a brief construct, the previous result that a strong SN is Turing-complete. Remind that, unlike Petri nets, an SN always fires enabled transitions at their maximal firing multiplicity, as a single step, leaving for a nondeterministic choice of which fireable transitions to fire. A strong SN restricts nondeterministic choice to firing only the transitions having the highest firing multiplicity. **Keywords:** Sleptsov net; Turing-completeness; place-transition net; multiple firing.

1. **Introduction**

The present paper proves that a *Sleptsov net* (SN) is Turing-complete, that considerably improves, with a brief construct, the previous result [1] that a *strong* SN is Turing-complete.

Remind that, unlike Petri nets, an SN always fires enabled transitions at their *maximal firing multiplicity*, as a single step, leaving for a nondeterministic choice of which fireable transitions to fire. A strong SN restricts nondeterministic choice to firing only the transition having the highest firing multiplicity.

The proof pattern follows [1], simulating a *Shepherdson and Sturgis register machine* (RM), proven to be Turing complete [2]. Remind that an RM implements three operations over a finite set of registers, each resister storing a nonnegative magnitude: increment, decrement (when a register is greater than zero), and zero check. Here we present an *SN that implements zero check* (fig. 1a) having same, as in [1], simple nets for increment (fig. 1b) and decrement (fig. 1c) implementation.

The results have been obtained and double-checked within modeling system Tina [3] which upcoming version supports SNs.

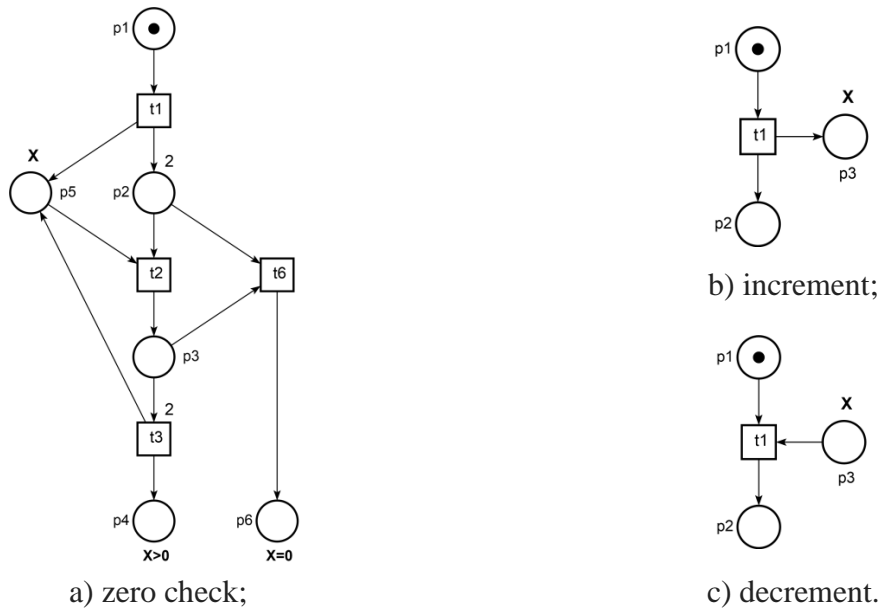

a) zero check;     b) increment;     c) decrement.

Fig. 1. SN components simulating instructions of RM.

## 2. Zero check with an SN

*Lemma 1.* SN in fig. 1a implements zero check of variable $X$.

*Proof.*

a) Suppose $X = 0$.
In $\bar{\mu} = \{p_1\}$, the only fireable sequence $\sigma = t_1 t_2 t_6$ fires:

$$\{p_1\} \xrightarrow{t_1} \{1 \cdot p_5, 2 \cdot p_2\} \xrightarrow{t_2} \{p_2, p_3\} \xrightarrow{t_6} \{p_6\}.$$

A token within place $p_6$ indicates that value of $X$ equals to zero.

b) Suppose $X > 0$.
In $\bar{\mu} = \{p_1, X \cdot p_5\}$, the only fireable sequence $\sigma = t_1 (2 \cdot t_2) t_3$ fires:

$$\{p_1, X \cdot p_5\} \xrightarrow{t_1} \{2 \cdot p_2, (X+1) \cdot p_5\} \xrightarrow{2 \cdot t_2} \{2 \cdot p_3, (X-1) \cdot p_5\} \xrightarrow{t_3} \{p_4, X \cdot p_5\}.$$

A token within place $p_4$ indicates that value $X$ is greater than zero.

Marking of other places, except of $\{p_1, p_4, p_6\}$, is not changed.

∎

Proof of Lemma 1 is illustrated with a parametric marking graph shown in fig. 2 and a trace of the net images shown in fig. 3, with highlighted in red firing transitions, obtained in system Tina [3]; in the line below a series of images, the number of firing transition copies is indicated in case it is greater than unit. We use letter $X$ to specify both a register of RM and its value; besides, we use local numbering of places and transitions within constructs of fig. 1, which are remunerated during composition of SN simulating an RM [1].

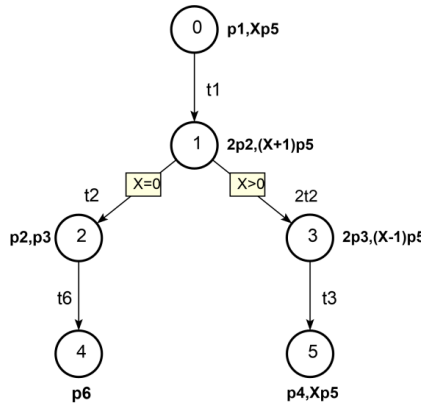

Fig. 2. Parametric marking graph for zero check net shown in fig. 1a.

*Theorem 1.* SN simulates RM.

Directly follows from Lemma 1 and RG simulation technique [1].

***Corollary.*** SN is Turing-complete.

Directly follows from Theorem 1 and [2].

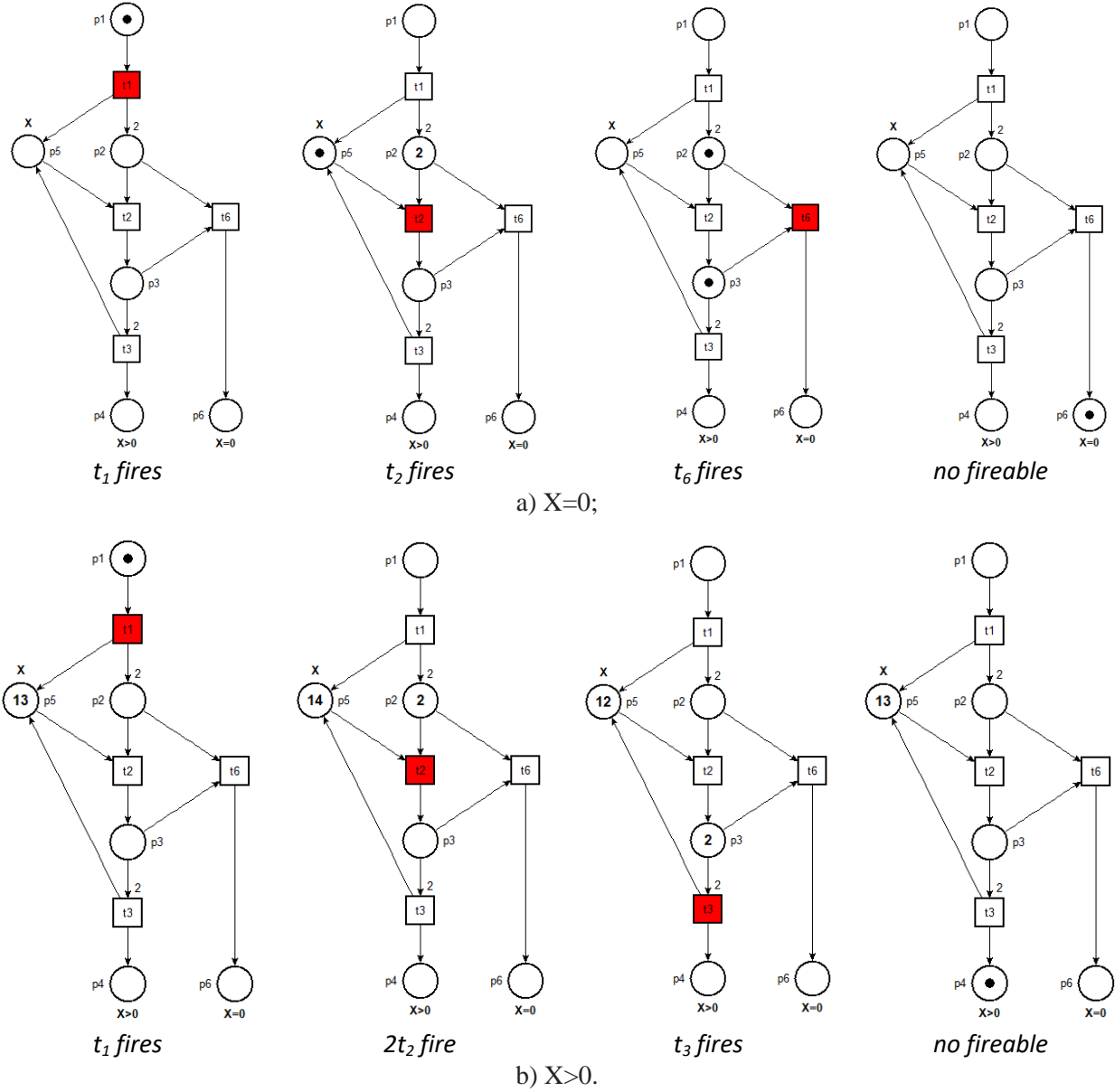

Fig. 3. Trace of transition firing sequences for zero check.

## 3. Conclusion

We have proven that an SN is Turing-complete i.e. capable of universal computations without any additions.